\documentclass{jps-cp}
\usepackage{txfonts} 

\usepackage{color}

\renewcommand{\vec}[1]{\mbox{\boldmath $#1$}}

\title{
Beyond the neutron-drip line: two-neutron decay of unbound nuclei}

\author{K. \textsc{Hagino}$^{1,2}$ and H. \textsc{Sagawa}$^{3,4}$}

\inst{$^{1}$
Department of Physics, Tohoku University, Sendai, 980-8578,  Japan \\
$^{2}$
Research Center for Electron Photon Science, Tohoku University, 1-2-1 Mikamine, Sendai 982-0826, Japan \\
$^{3}$
RIKEN Nishina Center, Wako 351-0198, Japan \\
$^{4}$
Center for Mathematics and Physics,  University of Aizu, 
Aizu-Wakamatsu, Fukushima 965-8560,  Japan}

\recdate{}

\abst{
We discuss a decay of unbound nuclei beyond the neutron 
drip-line using a three-body model with a core nucleus and two valence 
neutrons. We particularly discuss the role of dineutron correlation between 
the valence neutrons in the two-neutron emission from the 
ground state of $^{26}$O and $^{10}$He nuclei. 
Our calculations clearly indicate that 
the emission 
of the two neutrons in the back-to-back direction is enhanced due to 
the dineutron correlation. 
}
\kword{neutron dripline, resonance, dineutron correlation, three-body model}

\begin{document}
\maketitle

\section{Introduction}

An important question in physics of unstable nuclei is: where are the 
neutron and the proton drip-lines located in the nuclear chart? 
On the neutron side, the drip-line has so far 
been identified experimentally 
up to oxygen isotopes \cite{Saku99}, which is being extended in new 
measurements e.g., at RIBF in RIKEN. 
In this connection, an interesting phenomenon has been found in the 
neutron drip-line for oxygen isotopes. That is, while the drip-line 
nucleus for oxygen is $^{24}$O (with the neutron number of $N$=16), 
the drip-line extends considerably for fluorine isotopes, for which 
at least up to $^{31}$F (with $N$=22) is bound \cite{Saku99}. 
This phenomenon has been referred to as oxygen anomaly \cite{Otsuka10}, 
for which Otsuka and his collaborators have successfully explained 
in terms of the three-body and the tensor 
interactions \cite{Otsuka10,Otsuka05}. 

Partly motivated by the oxygen anomaly, in this contribution we 
discuss the decay dynamics of the unbound $^{26}$O nucleus, which is 
located beyond the neutron drip-line. The decay of $^{26}$O has attracted 
lots of attention in recent years both experimentally 
\cite{LDK12,CSA13,KBB13,KBC15,Kondo15} 
and theoretically 
\cite{GMSZ11,GMZ13,GZ15,HS14,HS14-2,HS16,Tsukiyama15,FRMN17,Wang17,Hove16,Desc17}. 
We shall use here a three-body model with an inert $^{24}$O core and two valence 
neutrons, and clarify the role of neutron-neutron correlation in the decay 
of $^{26}$O \cite{HS14,HS14-2,HS16}. We shall apply a similar model also 
to $^{10}$He and discuss the angular correlation of the two 
emitted neutrons from a spontaneous decay of the unbound $^{10}$He nucleus.  

\section{Two-neutron decay of $^{26}$O} 

\subsection{The decay energy spectrum} 

We first discuss the unbound $^{26}$O nucleus. 
There have been three measurements for the decay energy spectrum of 
$^{26}$O, at Michigan State University (MSU) \cite{LDK12}, GSI \cite{CSA13}, 
and RIKEN \cite{Kondo15}. In all of these three measurements, a proton 
knockout reaction of $^{27}$F has been used to produce $^{26}$O, 
which then spontaneously decays to $^{24}$O + $n$ +$n$. 
The decay energy of $^{26}$O has been determined in the RIKEN measurement 
to be 18$\pm 3$ (stat) $\pm 4$ (syst) keV \cite{Kondo15}. 
The $^{26}$O nucleus was therefore almost bound if the nuclear 
interaction was a little bit stronger. 
In reality, the interaction is not strong enough so that the ground state 
of $^{26}$O appears as a resonance state. 

To describe the two-neutron decay of $^{26}$O, we employ a three-body model. 
The Hamiltonian in this model reads, 
\begin{equation}
H=h_{nC}(1)+h_{nC}(2)+v(\vec{r}_1,\vec{r}_2), 
\end{equation}
where $h_{nC}$ is a single-particle Hamiltonian 
for the relative motion between a neutron and the core nucleus, and 
$v(\vec{r}_1,\vec{r}_2)$ is the interaction between the valence neutrons. 
For simplicity, we have neglected the off-diagonal component of the 
recoil kinetic energy of the core nucleus \cite{HS14,HS14-2,HS16}. 
For the interaction $v$, we take a contact interaction whose strength 
depends on the density \cite{BE91,HS05,SH15}, that is, 
\begin{equation}
v(\vec{r}_1,\vec{r}_2)=\delta(\vec{r}_1-\vec{r}_2)
\left(v_0+\frac{v_\rho}{1+\exp[(r_1-R_\rho)/a_\rho]}\right), 
\label{vnn}
\end{equation}
where the strength $v_0$ for the density independent part is determined 
from the scattering length for $nn$ scattering\cite{EBH97}, 
while the parameters for the 
density dependent part, $v_\rho$, $R_\rho$, and $a_\rho$, 
are adjusted 
to reproduce the 
ground state energy. 

\begin{figure}[t]
\begin{center}
\includegraphics[scale=0.6,clip]{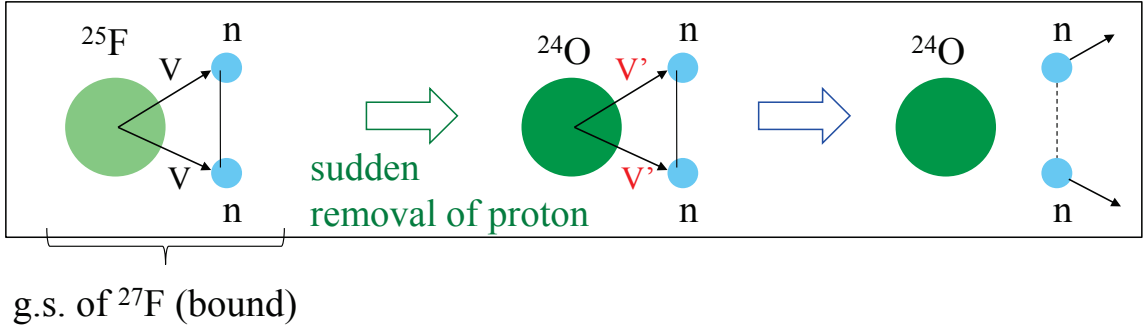}
\end{center}
\caption{
A schematic illustration of the three-body model calculation for the 
two-neutron 
decay of $^{26}$O. }
\end{figure}

Figure 1 illustrates schematically how the three-body model can be applied to 
the two-neutron decay of $^{26}$O nucleus \cite{HS14,HS16}. 
Since the $^{26}$O nucleus is produced experimentally 
from $^{27}$F, we first apply 
the three-body model to the $^{27}$F nucleus 
and construct the bound ground state wave function of this nucleus as 
$\Psi_{nn}(^{27}{\rm F})
\otimes|^{25}{\rm F}\rangle$, 
where $\Psi_{nn}(^{27}{\rm F})$ is the wave function for the two 
valence neutrons 
while $|^{25}{\rm F}\rangle$ is the ground state wave function for the 
core nucleus, $^{25}$F. 
We then assume a sudden removal of proton from the core nucleus. 
That is, the core nucleus suddenly changes from $^{25}$F to 
$^{24}$O while the two neutron configuration, $\Psi_{nn}(^{27}{\rm F})$, 
remain the same, 
therefore the three-body wave function becomes 
$\Psi_{nn}(^{27}{\rm F})\otimes|^{24}{\rm O}\rangle$. 
The sudden removal of proton also 
accompanies a change in the single-particle Hamiltonian, $h_{nC}$. 
Since the wave function 
$\Psi_{nn}(^{27}{\rm F})\otimes|^{24}{\rm O}\rangle$ is not 
an eigenfunction of the three-body model Hamiltonian for $^{26}$O, it 
forms a wave packet. 
Because $^{26}$O does not have a bound state, 
the wave function then spontaneously evolves in time and 
the two valence neutrons fly away from the core nucleus, $^{24}$O. 
The decay energy spectrum can be constructed by taking an overlap 
between the initial wave function, $\Psi_{nn}(^{27}{\rm F})$, and 
the two-particle wave function for $^{26}$O, 
$\Psi_{nn}(^{26}{\rm O};E)$, as 
\begin{equation}
\frac{dP}{dE}=|\langle\Psi_{nn}(^{27}{\rm F})|
\Psi_{nn}(^{26}{\rm O}; E)
\rangle|^2.
\label{dPdE}
\end{equation}
We mention that a very similar idea has been put forward also by Tsukiyama, 
Otsuka, and Fujimoto, who studied the decay of $^{25}$O and $^{26}$O using 
a shell model \cite{Tsukiyama15}. 

The decay energy spectrum, Eq. (\ref{dPdE}), can be expressed also in 
terms of the Green's function as, 
\begin{equation}
\frac{dP}{dE}=
\int dE'|\langle\Psi_{E'}|\Phi_{\rm ref}\rangle|^2
\,\delta(E-E')=
\frac{1}{\pi}\Im
\left\langle\Phi_{\rm ref}\left|\,\frac{1}{H-E-i\eta}\,\right|\Phi_{\rm ref}
\right\rangle, 
\end{equation}
where $|\Phi_{\rm ref}\rangle\equiv |\Psi_{nn}(^{27}{\rm F})\rangle$ 
is the initial (reference) state, and $\Im$ denotes the imaginary 
part. Notice that $1/(H-E-i\eta)$, is nothing but the Green's function, 
$G(E)$, with $\eta$ being an infinitesimally small positive number. 
The Green's function can be constructed with the uncorrelated Green's 
function, $G_0(E)=1/(h_{nC}(1)+h_{nC}(2)-E-i\eta)$, as \cite{EB92}, 
\begin{equation}
G(E) = G_0(E)-G_0(E)v(1+G_0(E)v)^{-1}G_0(E). 
\end{equation}
Since we employ a contact interaction for $v$, this equation can 
be solved most easily in the coordinate space \cite{BE91,EB92}. 

\begin{figure}[t]
\begin{center}
\includegraphics[scale=0.6,clip]{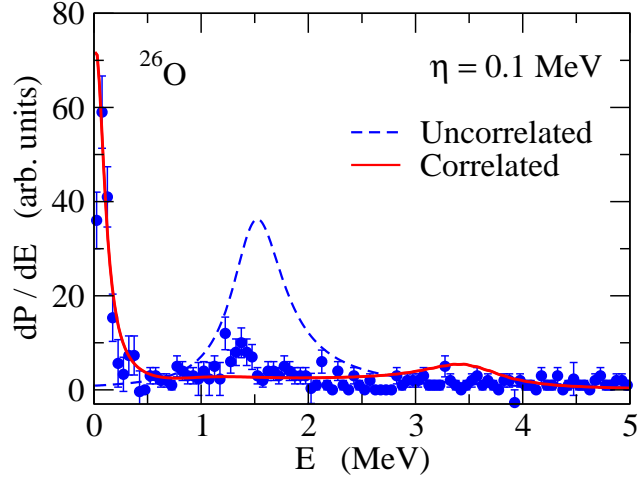}
\end{center}
\caption{
The decay energy spectrum for $^{26}$O obtained with the three-body 
model. The dashed line shows the result for the uncorrelated case, in 
which the $nn$ interaction is set to zero. The solid line 
shows the result for the full correlated case. The experimental data 
are taken from Ref. \cite{Kondo15}. }
\end{figure}

Figure 2 shows the decay energy spectrum of $^{26}$O so obtained. 
To this end, we employ a simple uncorrelated (1$d_{3/2})^2$ configuration in 
$^{27}$F for the initial reference state, $|\Phi_{\rm ref}\rangle$. 
We also keep $\eta$ to be a finite value, $\eta=0.1$ MeV, 
for a presentation purpose. 
When the $nn$ interaction is set to zero in the three-body Hamiltonian 
for $^{26}$O, 
the initial reference state, $|\Phi_{\rm ref}\rangle$, 
has a large overlap with the 
(1$d_{3/2})^2$ configuration in $^{26}$O, for which the 
1$d_{3/2}$ resonance state in $^{25}$O 
is located at 749 keV \cite{Kondo15}. 
The decay energy spectrum then has a peak at twice this energy, 1.498 MeV 
(see the dashed line). 
When the $nn$ interaction is switched on, the peak is shifted largely towards 
lower energies. If the parameters for the $nn$ interaction are adjusted to 
the ground state energy of $^{26}$O, the experimental decay energy spectrum 
is well reproduced with this calculations, as is indicated by the 
solid line in the figure. 

\subsection{The 2$^+$ state}

The experimental data shown in Fig. 2 
show a prominent secondary peak at $E=1.28^{+0.11}_{-0.08}$ 
MeV \cite{Kondo15}. 
This is most likely the 2$_1^+$ state in $^{26}$O, while the dominant peak 
at 18 keV is due to the 0$_1^+$ state. 
Our three-body model calculation yields the 2$^+$ peak at 1.282 MeV with 
the width of $\Gamma = 0.12$ MeV \cite{HS16}, which agrees perfectly with 
the experimental data. An important fact here is that the energy shift 
from the uncorrelated case, 1.498 MeV, is much smaller for the 2$^+$ state 
as compared to the ground state 0$^+$ state. This is typically the case for 
a single-$j$ configuration with a pairing interaction \cite{RS80}, 
in which the overlap 
between the wave functions for the two neutrons in 2$^+$ state is much 
smaller than that in 0$^+$ state, and thus the energy gain due to the 
pairing interaction is much smaller. 
Of course, the dineutron correlation, that is, a mixture of several 
configurations, also plays an important role, especially for the 0$^+$ state, 
but the relative position of the 0$^+$ and 2$^+$ peaks, with respect to the 
uncorrelated energy, can be qualitatively understood in this way. 
We mention that a similar result has been obtained recently 
also with a Gamow shell model calculation \cite{Wang17}. 

\subsection{The angular correlation}

Let us now discuss the angular correlation between the emitted two 
valence neutrons. Very roughly speaking, the angular correlation 
can be computed as, 
\begin{equation}
P(\theta_{nn})\sim 4\pi\int dk_1dk_2|\langle k_1k_2,\theta_{nn}
|\Psi_{nn}(E)\rangle|^2, 
\end{equation}
where $|k_1k_2,\theta\rangle$ is an uncorrelated wave function 
for the valence neutrons with the asymptotic wave numbers of 
$\vec{k}_1=(k_1,\hat{\vec{k}}_1=0)$ and 
$\vec{k}_2=(k_2,\hat{\vec{k}}_2=\theta_{nn})$. 
$|\Psi_{nn}(E)\rangle$ is a three-body wave function 
obtained with the three-body 
model 
at 
$E=(k_1^2\hbar^2+k_2^2\hbar^2)/(2\mu)$, where $\mu$ is the reduced 
mass between a valence neutron and the core nucleus.  
See Refs. \cite{HS14,HS16,EB92} for a more accurate formula with 
phase shifts and the Green's function. 

\begin{figure}[t]
\begin{center}
\includegraphics[scale=0.6,clip]{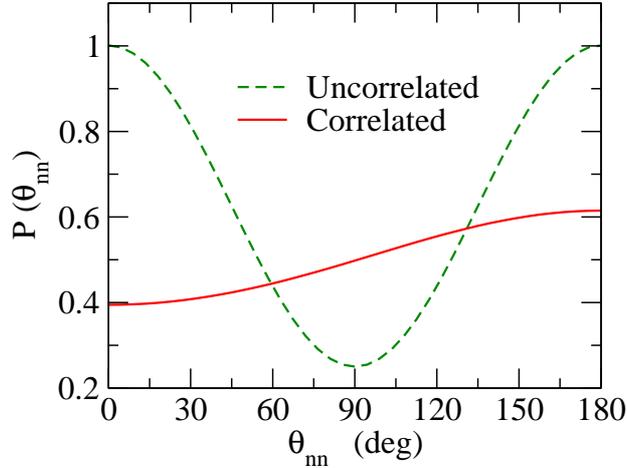}
\end{center}
\caption{
The angular correlation in the momentum space 
for the two valence neutrons emitted from the decay of $^{26}$O. 
The dashed and the solid curves show the results of the uncorrelated 
and the correlated cases, respectively. }
\end{figure}

Figure 3 shows the angular distribution for the decay of $^{26}$O. 
The dashed and the solid lines show the results for the uncorrelated 
and the correlated cases, respectively. 
In the uncorrelated case, the angular correlation is symmetric with 
respect to $\theta_{nn}=\pi/2$. On the other hand, for the 
correlated case, the component for large angles is enhanced while the 
small angle component is significantly 
suppressed. The angular distribution then becomes 
asymmetric, with an enhancement of back-to-back emissions. 
This is as a natural 
consequence of the dineutron correlation \cite{HS-FewBody}, 
with which the 
valence neutrons are spatially localized at a similar position 
inside a nucleus. 
The localization in the coordinate space corresponds 
to a large relative momentum for the two neutrons \cite{HS16}, 
which can be understood from a view of 
uncertainty relation. 
Once the two neutrons are emitted outside the core nucleus, 
the two neutrons then move in the opposite direction. 
This clearly indicates that an observation of the enhancement of 
back-to-back emission makes a direct experimental evidence for the 
dineutron correlation. 
Another three-body model calculation by Grigorenko 
{\it et al.} has also yielded a similar enhancement of 
back-to-back emission \cite{GMZ13}. 

\section{Two-neutron decay of $^{10}$He} 

Let us next discuss the two-neutron decay of 
$^{10}$He. 
In the experiments reported in Ref. \cite{Johansson10}, 
the $^{10}$He nucleus was produced by a proton knockout reaction 
from $^{11}$Li. The experimental decay energy spectrum shows a 
prominent peak at 1.54(11) MeV with a width of 
$\Gamma$ = 1.91 (41) MeV \cite{Johansson10}, 
which has been confirmed recently with a 2$p$2$n$-removal 
reaction from $^{14}$Be \cite{Kohley12}. 
See also Ref. \cite{Korsheninnikov94} for an earlier measurement. 

On the other hand, 
the structure of $^9$He nucleus, which is used to calibrate 
the parameters for the three-body model for $^{10}$He, 
has been poorly understood. 
The authors of Ref. \cite{Johansson10} fitted 
the observed decay energy spectrum with a narrow $1/2^-$ resonance 
at 1.33(8) MeV, whose width 
has been deduced to be $\Gamma$=0.10(6) 
MeV using $^9$Be($^{14}$C,$^{14}$O)$^9$He reaction \cite{Bohlen99}. 
This resonance state has been identified at 2.0 $\pm$ 0.2 MeV with 
a width of $\sim$ 2 MeV, and at 1.235$\pm$ 0.115 MeV with a width of 
130$^{+170}_{-130}$ keV in Refs. \cite{Golovkov07} and \cite{AlKalanee13}, 
respectively. 
In contrast, a recent high-resolution measurement of 
$^8$He+$p$ scattering does not support 
a narrow $1/2^-$ resonance in $^9$He\cite{Uberseder16}. 
In addition to the  
$1/2^-$ resonance, many experimental data also show 
a large strength close to the threshold, which is interpreted as 
an $s$-wave virtual state. 
The experimental data have been fitted with 
the $s$-wave scattering length 
for $n$-$^8$He scattering 
of $a_s=-3.17 (66)$ \cite{Johansson10}, 
$a_s> -20$ fm \cite{Golovkov07}, $a_s=-12\pm3$ fm \cite{AlKalanee13}, 
and $a_s < -10$ fm \cite{Chen01}, among which the value 
in Ref. \cite{Johansson10} seems incompatible with the others. 
On the other hand, the measurement reported in 
Ref. \cite{Uberseder16} has suggested a broad 1/2$^+$ state 
at approximately 3 MeV above the threshold, rather than 
near the threshold. 

\begin{figure}[t]
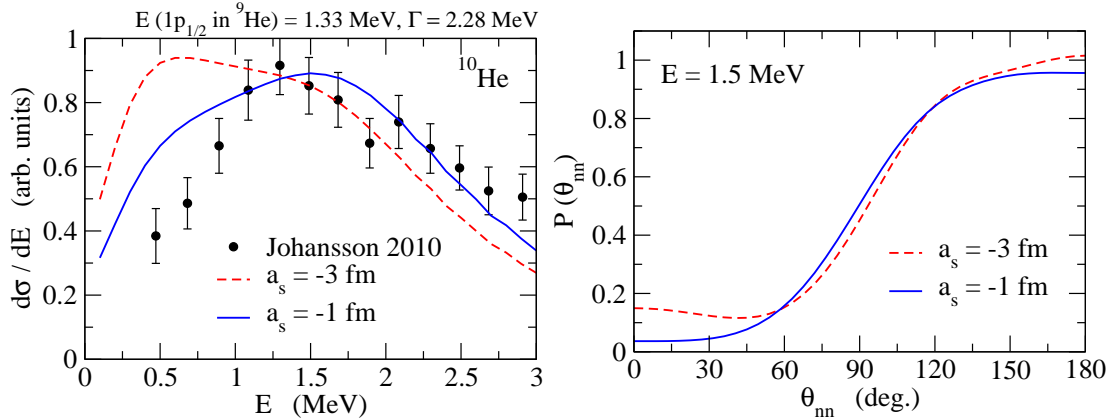

\begin{center}
\includegraphics[scale=0.5,clip]{fig4.eps}
\includegraphics[scale=0.5,clip]{fig5.eps}
\caption{
(Left panel) The decay energy spectrum for the $^{10}$He nucleus 
obtained with a three-body model of $^8$He+$n$+$n$. 
The dashed and the solid lines are obtained with the $s$-wave 
scattering length for the $n$-$^8$He scattering of $a_s=-3$ fm 
and $-1$ fm, respectively. The experimental data are taken 
from Ref. \cite{Johansson10}. 
(Right panel) The angular correlation for the emitted two neutrons 
from the two-neutron decay of $^{10}$He at the decay energy 
of $E$=1.5 MeV. The meaning of each line 
is the same as in the left panel.}
\end{center}
\end{figure}

Even though there is a large uncertainty in the structure of 
the $^{9}$He nucleus,
we perform a three-body model calculation for the two-neutron 
decay of $^{10}$He assuming a $^8$He+$n$+$n$ structure. 
To this end, we assume a simple 
$(1s_{1/2})^2(1p_{3/2})^4$ 
structure for the neutron configuration of the core nucleus, $^8$He, 
even though an actual structure may be much more complex \cite{HTS08}. 
For the neutron-core potential, we introduce a parity-dependent 
Woods-Saxon potential. For negative parity states, 
we adjust the parameters 
to reproduce the $p_{1/2}$ resonance at 1.33 MeV. 
Such potential, however, yields a much larger resonance 
width, with an order of MeV, as compared to the 
width of around 0.1 MeV observed in Refs. 
\cite{Bohlen99,AlKalanee13}. 
This may be due to either an artifact of 
our assumption of a simple structure for the 
$^8$He nucleus or the experimental uncertainty of the $p_{1/2}$ resonance 
state, as the calculated width is consistent with the observed width 
reported in Ref. \cite{Golovkov07}.
For even partial waves, we vary 
the depth parameter of the Woods-Saxon potential 
in order to investigate the dependence of the results on the 
the $s$-wave scattering 
length, $a_s$. In the three-body model calculations, we include 
partial waves up to $\ell$ = 14 with an energy cut-off of 
30 MeV. 

The dashed line in the left panel of Fig. 4 shows the 
decay energy spectrum of $^{10}$He 
obtained with $a_s=-3$ fm, which is close to the empirical value 
of Ref. \cite{Johansson10}. In this calculation, 
we use the ground state wave function 
for $^{11}$Li obtained with a three-body model \cite{HS05} as a reference 
state, $\Phi_{\rm ref}$, and also 
a similar density-dependent contact interaction 
as the one used in 
Ref. \cite{HS05} 
for the interaction between the valence neutrons in $^{10}$He. 
As one can see in the figure, 
this calculation does not reproduce well the observed 
decay energy spectrum. Again, this may be due to the simple structure assumed 
for the core nucleus, $^8$He. 
By changing the value of $s$-wave scattering length from $a_s=-3$ fm to 
$a_s=-1$ fm, one could make the calculation 
agree better with the experimental data, as is indicated by the 
solid line in the figure, even though this value of scattering length 
is inconsistent with the experimental values.

The angular correlation for the emitted two neutrons 
is plotted in the 
right panel of Fig. 4 for the decay energy of $E$=1.5 MeV. 
As in the $^{26}$O nucleus shown in Fig. 3, 
the back-to-back 
emission is considerably enhanced due to the dineutron correlation. 
It is interesting to notice that the angular correlation is less sensitive 
to the value of the $s$-wave scattering length as compared to the decay 
energy spectrum. 
Evidently, 
the angular correlation provides 
a good probe for the dineutron correlation, which has 
otherwise been difficult 
to observe experimentally. 

\section{Summary and future perspectives}

We have discussed the two-neutron emission decay of unbound 
nuclei beyond the neutron drip-line using the 
three-body model with a density dependent contact interaction 
between the valence neutrons. 
We have first calculated the decay energy spectrum for $^{26}$O 
using a method based on 
the two-particle Green's function, which is easy to evaluate 
in the coordinate space with a contact interaction. 
We have shown that a peak in the energy spectrum is considerably 
shifted towards low energies due to the interaction between the 
valence neutrons. By adjusting the parameters for the interaction 
so that the ground state energy is reproduced, we have shown 
that one can achieve an excellent agreement 
with the experimental data for the excited 2$^+$ state. 
We have then investigated the angular correlation of the 
two emitted neutrons, and have shown that 
an emission of 
the two neutrons in the opposite direction (that is, 
the back-to-back emission) 
is enhanced due to the dineutron correlation. 
We have applied the same model also to the $^{10}$He nucleus assuming 
a simple $p_{3/2}$-closed structure for the core nucleus, $^8$He. 
As in $^{26}$O, we have found that the back-to-back two-neutron 
emission is preferred in the decay of $^{10}$He. 

As we have discussed in this contribution, 
two-neutron decays of unbound nuclei provide an important probe for the 
dineutron correlation inside nuclei. 
The dineutron correlation has been predicted theoretically for some time, 
but it has not been straightforward to probe it experimentally. 
In the Coulomb dissociation of Borromean nuclei, 
one can use the cluster sum rule 
to deduce the mean value of 
the opening angle between the valence neutrons \cite{N06,HS07,BH07}. 
The extracted mean opening angles for $^{11}$Li and $^6$He 
are significantly smaller than 
the value for the uncorrelated case \cite{N06,HS07,BH07}, 
that is, 
$\langle\theta_{nn}\rangle$=90 degrees, clearly indicating the existence 
of the dineutron correlation in these Borromean nuclei. 
A small problem in this analysis is, however, that one can access only to 
the mean value of a distribution and the detailed distribution 
cannot be probed. 

The two-proton radioactivity, that is, a spontaneous emission of 
two valence protons, of proton-rich nuclei \cite{PKGR12} 
has been expected to provide an alternative tool to probe the nucleon-nucleon 
correlation in the initial wave function. 
An attractive feature of this phenomenon is that the two valence protons 
are emitted directly  
from the ground state even without any external perturbation. 
The long range interaction between the two protons, however, make 
theoretical analyses quite complicated, and it may not be straightforward 
to probe the diproton correlation from this phenomenon. 

The two-neutron decay of unbound nuclei beyond the neutron-drip 
line is an analogous process of the two-proton radioactivity, 
corresponding to a penetration of two neutrons over a centrifugal 
barrier. Since the long range Coulomb interaction is absent, 
the dineutron correlation may be better probed using the two-neutron 
decays. In this context, we have argued in this contribution that 
an enhancement of back-to-back emission makes a direct evidence for the 
dineutron correlation. 
Its observation 
has remained as an experimental challenge, especially for the 
$^{26}$O nucleus, for which the decay energy is considerably small. 
Another experimental challenge is to measure the spin of the emitted 
two neutrons, which would carry an important information on the 
$nn$ correlation inside a nucleus. 

A theoretical challenge, on the other hand, is to 
extend the three-body model by including the core deformation. 
This will be important in discussing the two-neutron decay of 
the $^{16}$Be nucleus \cite{Spyrou12}. 
Another important challenge is an extension from the 
three-body to multi-body 
descriptions. For the two-neutron decay of $^{13}$Li 
\cite{Johansson10,Kohley13} and $^{10}$He, 
the nuclei $^{11}$Li and $^{8}$He would not make a good core, 
and such extension will be important. 
That is, a five-body description for $^{13}$Li with $^9$Li+4$n$ and 
a seven-body description for $^{10}$He with $^4$He+6$n$ would provide 
better results. 
One will also be able to apply a 
five-body model to four-neutron emission decays, such as a decay of $^{28}$O 
nucleus. A measurement for the decay of $^{28}$O has already 
been carried out in RIKEN 
and the experimental data have now been under 
analysis \cite{Kondo18}. It will therefore be a very important future 
direction to develop a theoretical model based on a 
five-body model which is applicable to the four-neutron 
emission decay of $^{28}$O. 

\section*{Acknowledgments}

We thank T. Nakamura and Y. Kondo for useful discussions. This work was 
supported by JSPS KAKENHI Grant Number 16H02179.

\end{document}